\documentstyle[aas2pp4,rotate]{article}

\slugcomment{Submitted 1997 April 17 to ApJ; Accepted 1997 June 23}

\lefthead{Ringwald et al.}
\righthead{PG~1002+506: a Be Star Apparently at Z $>$ +10 kpc}

\begin{document}

\title{PG~1002+506: a Be Star Apparently at Z $>$ +10 kpc}

\author{F. A. Ringwald}
\affil{Department of Astronomy and Astrophysics, The Pennsylvania State
University, \\ 525 Davey Laboratory, University Park, PA 16802-6305}
\authoremail{ringwald@astro.psu.edu}

\author{W. R. J. Rolleston}
\affil{Department of Pure \& Applied Physics, Queen's University of
Belfast, \\ Belfast BT7~1NN, Northern Ireland}
\authoremail{R.Rolleston@Queens-Belfast.ac.uk}

\author{R. A. Saffer}
\affil{Department of Astronomy and Astrophysics, Villanova University, \\
800 Lancaster Ave., Villanova, PA 19085}
\authoremail{rsaffer@ucis.vill.edu}

\and

\author{John R. Thorstensen}
\affil{Department of Physics and Astronomy, Dartmouth College, Hanover, 
NH 03755-3528} 
\authoremail{thorstensen@dartmouth.edu}

\begin{abstract}

PG~1002+506 is found to be a Be star, one of two so far found by the
Palomar-Green survey.  Its spectrum is classified as a B$5\pm 1$ Ve, with
$T_{\rm eff} = 14,900 \pm 1200,$ $\log g = 4.2 \pm 0.2,$ and $v \sin i =
340 \pm 50$~km~s$^{-1}.$ At $b = +51^{\circ},$ its height above the
Galactic plane would therefore be $Z = +10.8$~kpc, putting this apparently
young, rapidly rotating star well into the Galactic halo.  Its
heliocentric radial velocity is found to be $-2 \pm 15$~km~s$^{-1},$
consistent with either having been formed in the Galactic disk and
subsequently ejected, or having been formed in the halo. 

\end{abstract}

\section{Introduction}

PG~1002+506 was discovered by the Palomar-Green UV-excess survey (Green,
Schmidt, \& Liebert 1986) and listed as a cataclysmic variable (CV). 
During a study of the CVs from this survey, Ringwald (1993) obtained
ultraviolet and red spectra, and tentatively reclassified it as a detached
subdwarf binary, noting H$\alpha$ in strong emission, unresolved at
10-\AA\ resolution.  Several puzzling aspects were noted, however,
including the near-constancy of the radial velocities throughout two
nights, consistent with no change other than that attributable to
atmospheric dispersion in an unrotated slit.  There was also no
significant variation in the equivalent width of H$\alpha,$ which one
might expect if this were a detached CV progenitor with the hot component
irradiating the facing hemisphere of its companion. 

That PG~1002+506 is not a CV was shown definitively by E. L. Robinson
(1995, private communication): it does not flicker, or have the erratic
variability ubiquitous in CVs.  This was found with high-speed
simultaneous {\it UBVR\/} photometry taken in 1995 June with the Stiening
photometer on the McDonald Observatory 2.1-m telescope.  In 25~min of
photometry with 1-s time resolution, all bands showed peak-to-peak
amplitudes of $<\,2$\%. 

This and further spectra have forced another reclassification of this
star, as a high-latitude Be star.  This is one of two known in the
Palomar-Green catalog, the other being PG~0914+001 (Saffer et al.~1997). 
An Oe star from this survey is also known, PG~2120+062 (Moehler, Heber, \&
Dreizler 1994). 

For reviews on Be stars, see Jaschek \& Jaschek (1987) and Slettebak
(1988).  About one in five non-supergiant B stars shows emission, mainly
in H$\alpha$ but sometimes also in H$\beta$ and higher Balmer lines. 
Struve (1931) attributed this to a disk, extruded by the star's rotation
near the breakup velocity, $ \sqrt{ G M / R}.$ What excites the emission
in Be stars is a long-standing mystery, however, as is their evolutionary
status.  Although Be stars often have an IR excess, PG~1002+506 is not an
IRAS source (Neugebauer et al.~1988).

\section{Blue spectrum}

A blue spectrum (Figure 1) was taken in service time with the Intermediate
Dispersion Spectrograph on the Isaac Newton Telescope on La Palma. This
1800-s spectrum was taken in photometric conditions in $2''$ seeing,
through a $1.73''$~slit, and has 1.5-\AA\ (FWHM) resolution.  The slit was
aligned to the parallactic angle, to avoid atmospheric dispersion effects;
the spectrum was taken when PG~1002+506 was nearly overhead, at an airmass
of 1.08. 

A spectral classification of B$5\pm 1$ V was arrived at by comparing this
spectrum to model atmospheres (Kurucz 1979) and published spectra (Jacoby,
Hunter, \& Christian 1984; Jaschek \& Jaschek 1987).  That this is a
main-sequence star and not a subdwarf is shown by the presence of the H13
and 14 lines.  That it is not a giant or supergiant is shown by the widths
of its Balmer lines, with FWZI of H$\gamma$ of $ 31 \pm 3$~\AA.  There is
no spectroscopic evidence that this star is a binary.

\section{Radial Velocity}

On 1997 January 3 UT, two 10-min exposures were obtained with the Modular
Spectrograph on the 2.4-m Hiltner Telescope at Michigan-Dartmouth-MIT
Observatory, Kitt Peak, Arizona.  The spectra covered from 4650 to 6727
\AA, and had 4-\AA\ (FWHM) resolution.  The weather was poor, with $>
1\arcsec$ seeing and rising humidity that forced a shutdown just after
these spectra were taken.  The spectrograph slit was set at the
parallactic angle, even though PG~1002+506 was only one hour east of the
meridian. The 1\arcsec\ slit projected to 3 \AA\ on the detector.  With
the mediocre seeing, we expect ``slit-painting'' velocity errors to be
small, probably $< 5$ km~s$^{-1},$ based on experience with similar sharp
lines in white dwarf/red dwarf binaries (Thorstensen, Vennes, \& Shambrook
1994).  The exposures were bracketed by HgNeXe exposures, for which the
RMS residual was $< 0.05$~\AA, and the maximum residuals for the weakest
lines were $< 10$~km~s$^{-1}.$ Most lines had residuals around
2~km~s$^{-1}.$

H$\alpha$ appears to be slightly resolved, and is in strong emission (see
Figure 2), with an equivalent width of $17.8 \pm 0.3 $~\AA\ and FWHM of
$580 \pm 30$ km~s$^{-1}.$ There is also emission in the core of H$\beta.$
By convolving H$\alpha$ with the derivative of a Gaussian with FWHM = 8
\AA\ and taking the zero of the convolution as the velocity (Schneider \&
Young 1980), we find heliocentric radial velocities of the spectra taken
at HJD 2450451.90425 and 2450451.91140 of +29.3 and +28.9~km~s$^{-1},$
respectively. The velocities of the O~I $\lambda$\,6300 \AA\ night sky
line were 1.6 and 0.7~km~s$^{-1},$ showing the accuracy of the wavelength
scale. 

However, the emission lines in Be stars are well known to be variable in
profile over timescales of days or longer, and are therefore not reliable
indicators of the systemic velocity.  The spectra were therefore summed
together and rectified, to remove continuum slope effects.  The radial
velocity was then measured from the absorption wings of H$\alpha$ by
convolving a positive and a negative Gaussian with the line profile and
taking the zero of this convolution as the velocity (Schneider \& Young
1980).  In all cases the Gaussians had 4 channels FWHM.  The separation
between the Gaussians was varied, from 24 to 20 to 16 \AA; the
corresponding heliocentric radial velocities are $-2.0,$ $-0.5,$ and
$-4.1$~km~s$^{-1}.$ Finding the line's centroid by fitting and subtracting
a linear approximation of the continuum, numerical integration of the
intensity, and taking the centroid (crudely, with the IRAF {\it splot\/}
`e' command) gave +0.4 km~s$^{-1}.$ We conclude that PG~1002+506 has a
heliocentric radial velocity of $-2 \pm 15$~km~s$^{-1}.$

\section{Model atmosphere analysis}

We have performed a model atmosphere analysis of the blue optical spectrum
to estimate the atmospheric parameters $T_{\rm eff}$ and log $g$, as well
as the projected stellar rotation velocity $v \sin i$. Our grid of
synthetic spectra was calculated with the radiative transfer code SYNSPEC
(Hubeny, Lanz, \& Jeffrey 1995), assuming the temperature and pressure
stratifications of Kurucz (1991).  The metal and helium abundances were
held fixed at the solar value. At the temperature and surface gravity of
spectral type B5V, the assumption of LTE is well justified. The
temperature and gravity grid points were $T_{\rm eff}$ = 13,000 --
17,000~K in steps of 1,000~K, and log $g$ = 3.5 -- 5.0 in steps of 0.5
dex. In addition each model was convolved with a rotational broadening
function at projected rotation velocities $v \sin i$ = 50 -- 350
km~s$^{-1}$ in steps of 50 km~s$^{-1}$ to produce a 3-dimensional fitting
grid. The stellar parameters were estimated by simultaneous variation
using a non-linear $\chi^2$ minimization algorithm. Details of the
synthetic spectrum calculations and the fitting algorithm are given by
Saffer et al.~(1994) and Saffer et al.~(1997). Due to the partial filling
in of the lower Balmer lines by emission from the circumstellar material,
we have restricted the analysis to the portion of the spectrum blueward of
H$\beta.$

The best-fit stellar parameters are $T_{\rm eff} = 14,900 \pm 1200$~K,
$\log g = 4.20 \pm 0.2$, and $v \sin i = 340 \pm 50$ km~s$^{-1}$ (see
Figure 1).  The quoted 1-$\sigma$ errors are based on counting statistics
and account for covariance for the fitting parameters; they also estimate
systematic errors.

\section{Evolutionary status}

The effective temperature, surface gravity, and very high rotational
velocity are fully consistent with a spectral classification of B5Ve.  The
breakup velocity expected for this star is 540 km~s$^{-1}.$ The fit places
this star in the area of confusion in the $T_{\rm eff}/\log g$ diagram
where the Population I main- sequence intersects the Population II blue
horizontal branch (BHB) (Sch\"onberner 1993; Bertelli et al.~1994).  For
example, PG~0832+676 at first appeared to be a young star far from the
Galactic plane, but turned out to be a nearby blue evolved star, upon
analysis of high-resolution spectra (Hambly et al.~1996).  However,
identification of PG~1002+506 as a BHB star is contradicted both by the
emission reversals in the H$\alpha$ and H$\beta$ absorption lines, and by
its high rotation velocity, since BHB stars are slow rotators (Peterson,
Rood, \& Crocker 1995). 

Assuming PG~1002+506 to be of Population I origin, we used the derived
atmospheric parameters and the evolutionary tracks of Claret \& Gimenez
(1992) to estimate the stellar mass and evolutionary age (see Table 1).  A
distance estimate was obtained from the absolute visual magnitude deduced
from the stellar mass, atmospheric parameters, and bolometric corrections
of Kurucz (1979).  PG~1002+506 has $B = 15.36$ (Green et al.~1986).
Assuming $B - V = -0.16$ for B5V stars (Allen 1973), and a reddening
$E(B-V) < 0.01$, inferred from the map of Burstein \& Heiles (1982), this
would imply a distance of 13.9 kpc, which for a Galactic latitude $b =
51^{\circ},$ corresponds to a z-distance of 10.8 kpc above the Galactic
plane. Although large, this is not unheard of (Kilkenny 1992).  For a
Galactic longitude $l = 165^{\circ},$ this would imply a galactocentric
radius of 17.1 kpc, putting PG~1002+506 at the outskirts of the Galaxy.

\section{Kinematical analysis}

As the existence of young objects at large distances from the star forming
regions of the Galactic disk is potentially interesting, we have performed
a kinematical analysis for PG~1002+506. Although no proper motion
information is available, it is possible to use the observed radial
velocity of a star to constrain its evolutionary history. A detailed
description of the method of analysis is given by Rolleston et al.~(1997). 

We first consider a scenario whereby PG~1002+506 has a zero velocity
component parallel to the Galactic disk, and ejection has occurred
perpendicular to the plane of the Galaxy. We have corrected the observed
heliocentric velocity for the effects of differential rotation (Fich et
al.~1989), assuming that the halo co-rotates with the disk, to determine
the stellar radial motion $(v_r)$ with respect to a standard of rest
defined by its local environment. Our initial assumption implies that the
observed radial velocity is a component of the stellar space motion
$(v_z)$ perpendicular to the disk. We then attempt to show that PG
1002+506 could have reached its present position in the Galactic halo
within its evolutionary lifetime, while reproducing the observed radial
velocity, and calculating the required ejection velocity. These
calculations have adopted the gravitational potential function of House \&
Kilkenny (1980). This analysis implicitly assumes that the star is ejected
from the disk shortly after birth, consistent with cluster ejection
simulations. 

The results of the kinematical analysis are given in Table 1. Given the
large z-distance, it is not surprising to find the ``time of flight'' to
be larger than the evolutionary age. We have therefore considered the
effects of errors in the derived atmospheric parameters and the radial
velocity measurement. By optimizing the values of $T_{\rm eff}$ and $\log
g$ such that they are self-consistent within the errors, it is possible to
increase the evolutionary age, so that it is greater than the predicted
flight time. For example, adopting values of $T_{\rm eff} = 13,750$~K and
$\log g = 4.0$ would imply an age of 115 Myr for a mass of 4.0
$M_{\odot}.$ Allowing an error of 15 km~s$^{-1}$ in the observed
heliocentric velocity also decreases the estimated flight time, but not
significantly, to 84 Myr.

\section{Conclusions}

PG~1002+506 appears to be a young, rapidly rotating B5Ve star at a
distance of 10.8 kpc from the Galactic plane, and at a galactocentric
radius of 17.1 kpc. The kinematical analysis suggests that it could have
attained its present Galactic position having been ejected from the disk
shortly after its formation. Furthermore, the required ejection velocity
of $\approx 230$~km~s$^{-1}$ can also be produced by the known mechanisms
predicted by Leonard (1993).  A detailed atmospheric analysis with
higher-quality spectra should still be done, to determine abundances and
confirm that PG~1002+506 really is a distant main-sequence star, and not a
nearby blue evolved star.  If PG~1002+506 really is 10.8 kpc from the
Galactic plane, interstellar absorption in this same spectrum would probe
a line through the Galactic halo otherwise difficult to acquire.

\acknowledgments\noindent
E.~Harlaftis took the blue spectrum with the Isaac Newton telescope, which
is operated on La Palma by the Royal Greenwich Observatory at the Spanish
Observatorio del Roque de los Muchachos of the Instituto de Astrofisica de
Canarias.  Michigan-Dartmouth-MIT Observatory is operated by a consortium
of the University of Michigan, Dartmouth College, and the Massachusetts
Institute of Technology.  Thanks also to Rob Robinson, Malcolm Coe,
Richard Green, Uli Heber, Gerrie Peters, and Richard Wade, for helpful
discussions.

\clearpage

\begin{center}
\begin{deluxetable}{ll}
\footnotesize
\tablewidth{150pt}
\tablecaption{Stellar parameters \label{tbl-1}}
\startdata
$T_{\rm eff}$ &    $14,900 \pm 1200$~K\\
$\log g$     &    $4.2 \pm 0.2$\\
$v \sin i$   &    $340 \pm 50$~km~s$^{-1}$\\
$v_{breakup}$&     540~km~s$^{-1}$\\ 
Mass         &     4.2 $M_{\odot}$\\
Age          &     50 Myr\\
             &            \\
$l$          &     165.072$^{\circ}$\\
$b$          &      50.943$^{\circ}$\\
$B$          &      15.36\\
Distance     &     13.9 kpc\\
$R_{galactocentric}$ &     17.1 kpc\\
             &            \\
z-distance    &   10.8 kpc\\
$v_{heliocentric}$  &   $-2 \pm 15$~km~s$^{-1}$\\
$v_z$         &     18.0~km~s$^{-1}$\\
$T_{flight}$  &         85 Myr\\
$v_{ej}$      &        229~km~s$^{-1}$\\
              &            \\
Age (OPT)     &     115 Myr\\
$v_z$ (OPT)   &     39.0~km~s$^{-1}$\\
$T_{flight}$ (OPT) &     84 Myr\\
$v_{ej}$ (OPT) &        249~km~s$^{-1}$\\
\enddata
\end{deluxetable}
\end{center}

\clearpage

\begin{figure}
\rotate[r]{
\plotone{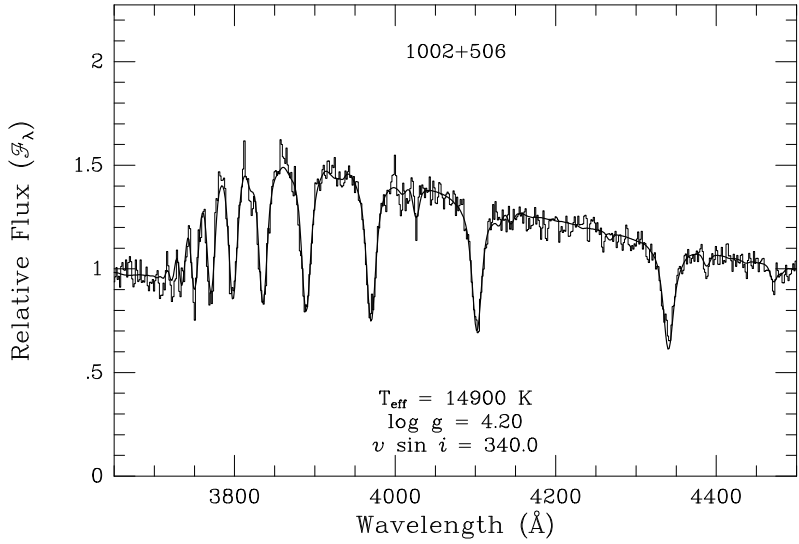}}
\caption{Spectrum of PG~1002+506, taken 1994 February 26 UT.  The
best-fit synthetic spectrum (heavy curve), simultaneously determining
$T_{\rm eff}$, $\log g$, and $v \sin i,$ is superimposed on the observed
spectrum (thin histogram). The core of H$\beta$ showed emission, but was
excluded to avoid spoiling the fit.\label{fig1}}
\end{figure}


\begin{figure}
\rotate[r]{
\plotone{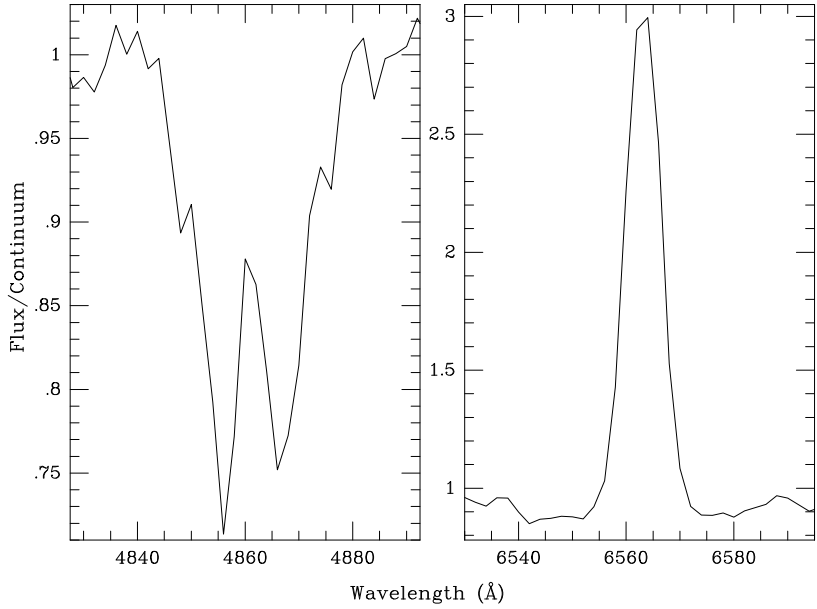}}
\caption{Modular Spectrograph profiles of H$\beta$ (left) and H$\alpha$
(right), at 4-\AA\ resolution, taken 1997 January 3 UT.\label{fig2}}
\end{figure}

\end{document}